\documentclass{aa}
\usepackage{graphicx}
\usepackage{natbib}
\usepackage{txfonts}
\bibpunct{(}{)}{;}{a}{}{,}    % to follow the A&A style

\def\apjs{ApJS}
        
\def\jgr{J. Geophys. Res.}
          
\def\grl{Geophys. Res. Lett.}

\def\aap{Astron. Astrophys.}

\def\apj{Astrophys. J.}

\def\solphys{Sol. Phys.}
     
\def\nat{Nature}

\begin{document}
\title{The place of the Sun among the Sun-like stars}
\author{A. I. Shapiro \inst{1}  \and W. Schmutz \inst{1} G. Cessateur\inst{1} \and E. Rozanov \inst{1,2}}
\offprints{A.I. Shapiro}

\institute{Physikalisch-Meteorologishes Observatorium Davos, World Radiation Center, 7260 Davos Dorf, Switzerland\\
\email{alexander.shapiro@pmodwrc.ch}
\and Institute for Atmospheric and Climate science ETH, Zurich, Switzerland\\}
\date{Received ; accepted }

\abstract
% context heading  (optional}
{Monitoring of the photometric and chromospheric HK emission data series of stars similar to the Sun in age and average activity level  showed that there is an empirical correlation between the average stellar  chromospheric activity level and the photometric variability.  In general, more active stars show larger photometric variability. Interestingly, the measurements and reconstructions of the solar irradiance show that the Sun is significantly less variable than indicated by the empirical relationship.
}
% aims heading (mandatory)
{We aim to identify possible reasons for the Sun to be currently outside of this relationship.}
% methods heading (mandatory)
{We employed different scenarios of solar HK emission and irradiance variability and compared them with available time series of Sun-like stars.} 
% results heading (mandatory)
{We show that the position of the Sun on the diagram of photometric variability versus chromospheric activity changes with time.  The present solar position is different from its temporal mean position as the satellite era of continuous solar irradiance measurements has accidentally coincided with a period of unusually high and stable solar activity. Our analysis suggests that although  present solar variability is significantly smaller than indicated by the stellar data, the temporal mean solar variability might be in agreement with the stellar data.  We propose that the continuation of the photometric program and its expansion to a larger stellar sample will ultimately allow us to constrain the historical solar variability.}
{}


\titlerunning{The place of the Sun among the Sun-like stars}
\maketitle

\section{Introduction}\label{sect:intro}
The comparison of the Sun with Sun-like stars helps to improve our understanding of both solar and stellar physics. One of its important applications is  the study of the solar irradiance variability. The consecutive measurements of the solar irradiance have only been available for the last three decades. The photometric observations of the large number of Sun-like stars can help to extend the time coverage and constrain the historical variability of the solar irradiance. 

The first monitoring of the photometric flux and chromospheric HK emission    \citep[which characterises the amount of non-thermal heating in the chromosphere and consequently  activity, see][]{HK} of the stars similar to the Sun in age and average activity level revealed an empirical correlation between the stellar photometric variability and mean chromospheric activity level  \citep{lockwoodetal1992}.  The continuation of the program \citep[see e.g.][]{lockwoodetal1997,radicketal1998,lockwoodetal2007} showed that the variability of a star with  solar chromospheric activity level  is expected to be about 0.1 \% in the 
Str{\"o}mgren filters b and y {  (centred at 467 and 547 nm, respectively)}. At the same time the regression of the observed total solar irradiance (TSI) variability to the variability in Str{\"o}mgren filters  b and y yielded a somewhat smaller value of about 0.04 \% \citep{radicketal1998}.

%showed that the solar brightness variations are significantly smaller than observed in the stellar survey \cite{lockwoodetal1992}. It was suggested that the present Sun could be in a unusually steady state of its evolution.  At the same time modern measurements of the solar irradiance and estimates using reconstructions give a significantly lower value for the present solar variability (0.02 - 0.04 \%) \citep{radicketal1998}.

Several attempts were made to explain this disagreement.  For example, \cite{foukal1994} proposed that the relatively low solar variability can be explained by  the dependence of the faculae to starspot ratio on the magnetic activity.
Later,  \cite{radick1994} showed that this explanation is not consistent with the stellar data.
\cite{schatten1993} suggested   that the special position of the Earth-based observer who sees the Sun from  its equatorial plane can reduce the variability of  solar brightness.  However, detailed calculations  showed that the  equatorial position of the observer leads to only a 30\% (most probable value) decrease in the solar variability and does not affect the mean level of the chromospheric activity \citep{knaacketal2003}. {  The out-of-ecliptic measurements of the Sun are needed to validate the model and accurately assess the effects from the hemispherical asymmetry of the density flux \citep{luis2012}.}

A significant part of the disagreement can also be attributed to the selection effect in the stellar sample. For example, \cite{Halletal2009}, using another stellar sample, presented a different version of variability versus activity regression. It is based on the monitoring of 28 Sun-like stars over time periods ranging from 6 to 15 yr. The slopes of the \cite{lockwoodetal2007}  and \cite{Halletal2009} regressions are approximately the same, but the \cite{Halletal2009} regression indicates about 0.1 dex less variability.

Recently \cite{analysis} proposed comparing the reconstructed time series of the solar irradiance with the photometry of Sun-like stars. In this paper we follow up on this idea. We re-evaluate the value of the solar variability in Str{\"o}mgren b and y filters  obtained by \cite{radicketal1998}  and show that the present location of the Sun on the variability versus activity diagram is anomalous. 

The continuous measurements of solar irradiance have been available only since the beginning of satellite observations (i.e. since 1978). Meanwhile, the proxy data indicate that the Sun has been  at a maximum plateau-state in its long-term evolution during the last 50 years \citep{lock_nature, solankietal2004}. Therefore, the amplitude of the 11-year activity cycle is expected to be relatively large, while the amplitude of the secular component, whose existence is actively speculated in the literature \citep[cf.][]{lockwood2011}, is very small in the most recent past. At the same time cosmogenic isotope records indicate that during the last millennia the level of the long-term solar activity has been significantly variable \citep{McCrackenetal2004,vonmoosetal2006}. This implies that the amplitude of the 11-year cycle also has been changing in time and, additionally, that the  Sun undergoes periods when the long-term photometric variations could be relatively strong.  Therefore, the temporal mean of  solar variability might be significantly different from the presently measured value.  To investigate this possibility we  locate the Sun on the variability versus chromospheric activity diagram,  employing different scenarios of solar long-term variability.

\section{Comparison of the stellar and present solar variability}\label{sect:present}
\begin{figure}
\resizebox{\hsize}{!}{\includegraphics{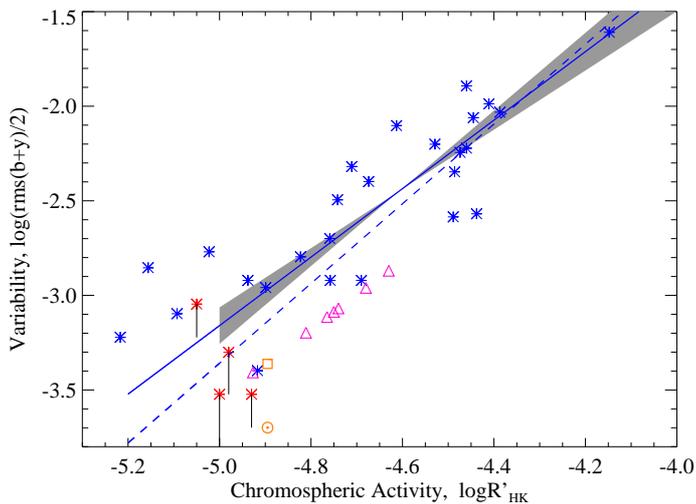}}
\caption{Photometric variability versus chromospheric activity. The blue stars  indicate the stars with observed variability from \cite{lockwoodetal2007}. Magenta triangles  indicate lower limits for stars with unconfirmed variability set at about 0.5 below the regression line. {  The solid (5.892+1.810\,${\rm log \, R'_{HK}} $) and dashed (7.180+2.108\,${\rm log \, R'_{HK}} $) lines} are regressions calculated excluding and including stars with unconfirmed variability, respectively. The shaded area corresponds to the  one-sigma error. The four red stars indicate HD10307, HD95128, HD168009, and HD146233 (18 Sco) from  \cite{Halletal2009}. Drop lines indicate the correction for  the variability of the comparison stars. The orange symbols denote the position of the Sun according to modern reconstructions (solar symbol) and the \cite{lockwoodetal2007} estimate (square symbol).}
\label{fig:diagr}
\end{figure}

Figure~\ref{fig:diagr}  presents the regression of the stellar variability versus chromospheric activity calculated with data from \cite{lockwoodetal2007}. To calculate the regression line we only consider the stars which were deemed variable by  \cite{lockwoodetal2007}. % and also excluded  HD201092.  The latter  was significantly above the regression line and only the upper limit of its variability was known  (as there was only one reliable comparison star). 

One can see that even for the stars with the same chromospheric activity level there is some divergence in the photometric variabilities. The variability of the less variable stars with low magnetic activity would be very difficult to detect (e.g. two of the five stars with unconfirmed variability had ${\rm log \, R'_{HK}} < -5$).  Therefore, the regression was made using only the stars with ${\rm log \, R'_{HK}} > -5$. 

Originally the Lowell Observatory program contained 41 stars \citep{lockwoodetal1997}. Some of the stars with unconfirmed variability  were later excluded from the program so  the final set contained only 32 stars. Seven stars from the original dataset had ${\rm log \, R'_{HK}} > -5$ but unconfirmed variability. The inability to detect the variability of the star is either because  its variability is indeed small or because the comparison stars are too variable. In the second case, the exclusion of the star is justified; however, in the former case the exclusion of such star from the analysis would affect the variability versus chromospheric activity dependency. To estimate the possible effect of such stars on the regression, we  located the stars {  (magenta triangles in Fig.~\ref{fig:diagr})} with unconfirmed variability below the regression line on the distance equal to the maximum deviation of the stars with confirmed variability from the regression line. While the original regression indicates that the root mean square (RMS) variability of a  star with the solar chromospheric activity level should be about  0.001 mag, the new regression yields a value of 0.00072 mag. These two numbers can be considered as lower and upper limits of the variability of a star with solar chromospheric activity level, yielded from the present stellar sample.

To take into account the new data of \cite{Halletal2009} we also calculated two regressions: one using the \cite{Halletal2009} data corrected for the variability of the comparison stars and another using   the \cite{Halletal2009} data combined with   the \cite{lockwoodetal2007} data. All these different versions of the regression yield solar variability {  within the error bars, estimated above (0.00072--0.001 mag)}. In Fig.~\ref{fig:diagr}  we show the variabilities of four stars from the ELODIE top ten solar analogs \citep{analogs} taken from  \cite{Halletal2009}. %One can see that two of these stars are located above our low estimate of the regression, while two are below (the variability of HD10307 is unconfirmed). 

Establishing the present solar position in Fig.~\ref{fig:diagr}  is not a straightforward task as there are no long-term measurements of the spectral solar variability in the band filters.   
\cite{radicketal1998} and \cite{lockwoodetal2007} assumed   the total solar irradiance and photometric variabilities to be connected as if they were caused by a change in the solar effective temperature.  This resulted in  0.00044  mag  estimation of the RMS of annual Str{\"o}mgren  (b+y)/2 values. 

The modern physically-based reconstructions of the solar irradiance that attribute the solar variability on the 11-year time scale to the competition between the dark sunspots and bright active regions  
\citep[e.g.][]{leanetal2005, krivova_rec2010, shapiro_rec},  give smaller values of the present solar variability for the last 50 years: 0.00027, 0.00017, and 0.00017 mag respectively.  We used the mean of these three numbers to locate the Sun in Fig.~\ref{fig:diagr}. 

Let us note that if instead of using the annual averages of the solar irradiance we emulate the stellar observations and consider the seasonal averages  then the calculated solar variability can be  increased by the rotational cycle.  
We calculated the real seasonal means using the  \cite{lockwoodetal2007} time series and estimated that the increase is generally smaller than 40\%.

The measurements obtained by spectral irradiance monitor (SIM) on-board of the solar radiation and climate experiment (SORCE) satellite  show yet another picture of solar variability over the last seven years \citep{harderetal2009}.  
The RMS of the annual (b+y)/2 values for 2004 -- 2011 period is 0.00029 mag. 
The extrapolation of the SIM annual (b+y)/2 values to the entire solar cycle using the annual values of the sunspot number results in 0.0004  mag  for the RMS.  \cite{fontenla2011} suggested that a physics-based extrapolation of the SIM data could put the Sun even closer to the regression line from Fig.~\ref{fig:diagr}.

\section{Comparison of the stellar and historical solar variability}\label{sect:historical}
Our main proposition is that the location of the Sun in Fig.~\ref{fig:diagr}  is not  fixed in time. Therefore, instead of using present solar variability for the comparison with Sun-like stars, one should use the temporal mean of the variability over a time interval long enough to reveal the entire range of  solar variability.      Because of the high solar activity the amplitude of the 11-year cycle has been relatively large during the last 50 years. So, if the 11-year cycle is the only contributor to the solar irradiance variability then the temporal mean of solar photometric variability is lower than its present value (see calculations below). This would imply that the disagreement between  the stellar and solar photometric variabilities is even more pronounced than presently thought. Instead, a secular component in  the solar irradiance can significantly increase the temporal mean of the solar photometric variability.

\begin{figure}
\resizebox{\hsize}{!}{\includegraphics{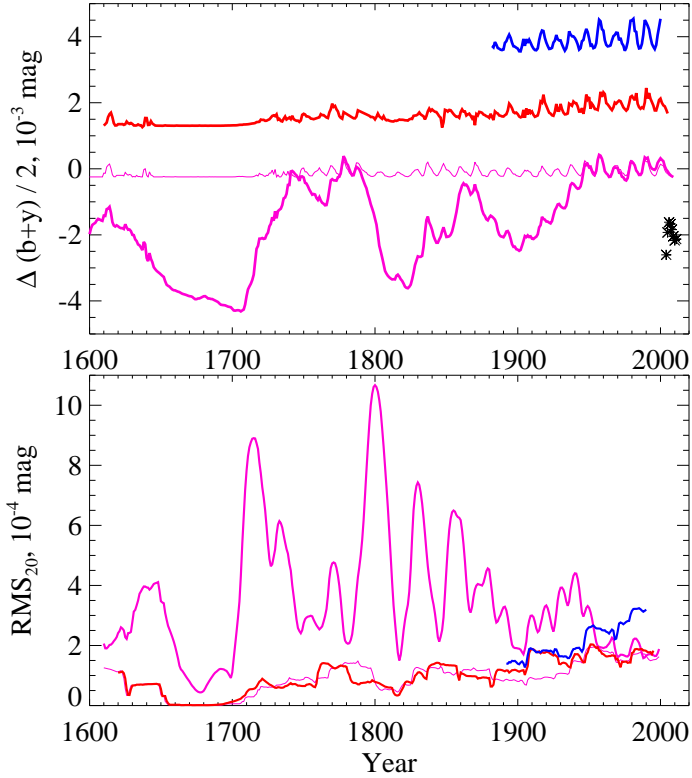}}
\caption{Upper panel: Annual (b+y)/2 values versus time according to the spectral solar irradiance (SSI) reconstructions to the past as calculated by  \cite{krivova_rec2010} (red), \cite{leanetal2005} (blue) and \cite{shapiro_rec} (magenta: thick and thin curves are original data and data with removed secular changes, respectively).  Stars indicates the SIM/SORCE measurements.
% Middle panel: rms  of the annual (b+y)/2 values versus time. The value for the year X is the rms of the set of annual (b+y)/2 values for the [X,2010] time period.
Lower panel:  Calculated for the 20 year time intervals RMS$_{20}$ values  versus time. 
%The value for the year X is the rms of the set of annual (b+y)/2 values for the [X-9,X+10] time period. Dash dot line indicates  the value rms$_{20}=4 \cdot 10^{-4}$ obtained from the SIM/SORCE observations. 
}
\label{fig:comp}
\end{figure}

The present consensus is  that secular changes in solar irradiance are proportional to secular changes in solar activity proxies. However, there is no agreement on the scaling coefficient \citep{leanetal2005,tappingetal2007,krivova_rec2010,schrijveretal2011,vieiraetal2011, shapiro_rec,Grec,analysis}.
To illustrate the effect of the secular variations on the solar photometric variability in Fig.~\ref{fig:comp} we plot  variations of annual (b+y)/2 values and magnitudes of  RMS$_{20}$ (the RMS$_{20}$  value for the year X is the RMS of (b+y)/2 values in the [X-9 yr,X+10 yr] dataset) for three recent spectral solar irradiance (SSI) reconstructions to the past. Most of the stars  used in this study
  had been observed for approximately 20 years, so the  RMS$_{20}$ value can be used for  direct comparison of the Sun and stars. According to all three reconstructions the secular changes in the solar irradiance have been very small during the last 50 years.  Therefore, during this period the solar variability (i.e. RMS$_{20}$ value) is roughly the same in all three reconstructions.
 At the same time, there were  periods before 1950 when the  long-term solar activity  changed significantly  \citep{McCrackenetal2004, steinhilberetal2008} and the reconstructions with large and small secular components are dramatically different.

%The positions of stars and the Sun in Fig.~\ref{fig:diagr} are calculated taking into account the noise of the measurements  \citep{lockwoodetal2007} and with  annually averaged data.
%Thus, the solar position is determined by the photometric variability due to the 11-year activity cycle  and the secular changes. 
%As was discussed above there is no consensus on the amplitude of both of these variabilities. 

 In order not to be limited by any particular reconstruction we introduce two parameters, $V_{11}$ and $V_{\rm LT}$, which describe the 11-year and long-term variability, respectively. We define ${V}_{11}$ as RMS variability of solar annual Str{\"o}mgren  (b+y)/2 fluxes  during cycles 21 and 22 (the 1976-1996 period) and $V_{\rm LT}$ as relative change of Str{\"o}mgren  (b+y)/2 flux between 22/23 and the Maunder minima. In Appendix~\ref{sect:model} we describe a simple model which allows us to calculate the solar variability over the millennia as a function of $V_{11}$ and $V_{\rm LT}$ parameters, using the cosmogenic isotope records \citep{McCrackenetal2004,vonmoosetal2006}. We show that while our approach does rely on the assumptions, they do not constrain its applicability, and by choosing corresponding $V_{\rm LT}$ and $V_{\rm 11}$ parameters we can roughly reproduce very different scenarios of the solar variability published in the recent literature. 
 
 Let us note that ${V}_{11}$ and $V_{\rm { LT}}$ values cannot be directly compared because ${V}_{11}$ corresponds to the RMS variation during a 20-year interval, whereas $V_{\rm { LT}}$  describes an absolute change over 300 years.

 An example for a chosen pair of parameters is illustrated in Fig.~\ref{fig:rec2400} (available as Online Material) for the time period from 400 BC to present.

The mean level of the solar chromospheric activity  can also change with time.  Currently there is no reason to suggest that it underwent significant changes in the past \citep{halllockwood2004,judgesaar2007}. 
\cite{saar2006} showed that the minimum chromospheric activity of stars with solar metallicity corresponds to $\log{R'_{HK}}=-5.08$.    We will now consider two extreme cases. The first one corresponds to the absence of a secular component in the mean solar chromospheric activity (which does not imply the absence of the secular component in the photometric variability).   So  changes of the solar activity index occur only because of  the evolution of the solar activity cycle. The second case (hereafter, case of strong secular changes) corresponds to the assumption that the mean solar chromospheric activity can reach a minimum  boundary value $\log{R'_{HK}}=-5.08$ during the periods of zero modulation potential. In  Appendix~\ref{sect:model} we show how the  mean solar chromospheric  activity can be reconstructed to the past in both of these cases.

%To reconstruct the  mean solar chromospheric  activity to the past we then scale it with modulation potential averaged over two solar cycles, assuming that zero value of the modulation potential corresponds to the case when the chromospheric activity  is equal to the present value during the solar minimum.
%  \cite[which corresponds to $\log{R'_{HK}}=-4.97$, see][]{judgesaar2007}. 
%The second case corresponds to the assumption that the mean solar chromospheric activity can reach a minimum  boundary value $\log{R'_{HK}}=-5.08$ during the periods of zero modulation potential (hereafter, strong secular changes).  

\begin{figure}
\resizebox{\hsize}{!}{\includegraphics{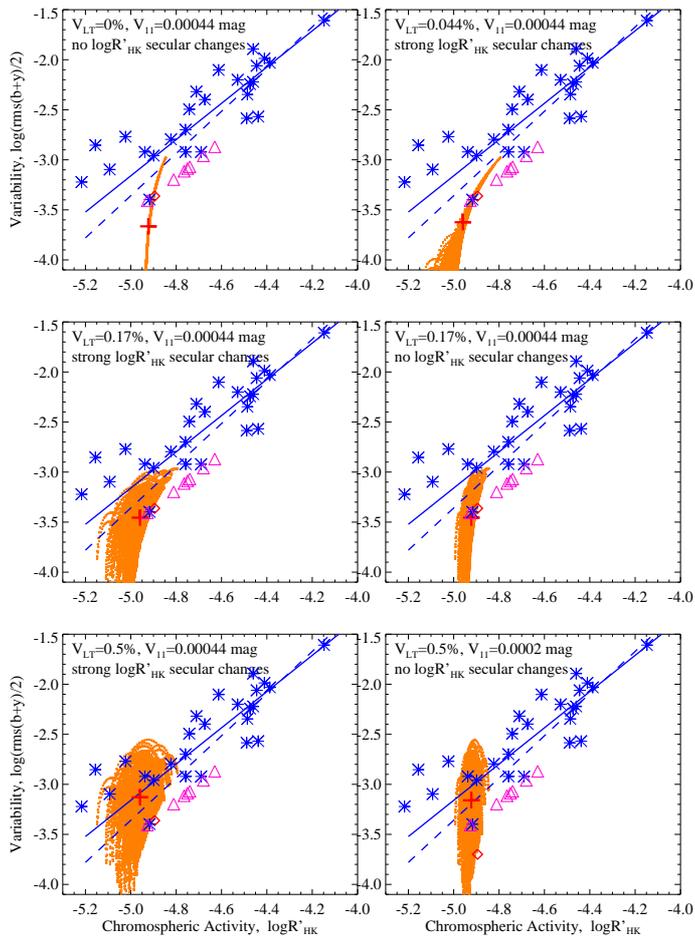}}
\caption{The same as Fig.~\ref{fig:diagr},  but with trajectories of the Sun over the last 9000 years (orange dots) for different sets of the long-term variability $V_{\rm {LT}}$ and the 11-year variability ${V}_{11}$. 
%Upper left panel:   $V_{\rm {LT}}=0$,  ${V}_{11}= 0.00044$ mag, no chromospheric long-term trend; upper right panel: $V_{\rm {LT}}=5.7 \cdot 10^{-3}$,  ${V}_{11}= 0.00044$ mag, no chromospheric long-term trend; lower left panel: $V_{\rm {LT}}=5.7 \cdot 10^{-3}$,  ${V}_{11}= 0.0002$ mag, no chromospheric long-term trend; lower right panel: $V_{\rm {LT}}=5.7 \cdot 10^{-3}$,  ${V}_{11}= 0.0002 $ mag, strongest possible chromospheric long-term trend.
%The shown areas of the variability {\it vs} chromospheric activity diagram correspond to the dashed rectangle on  Fig.~\ref{fig:diagr}.  
The red square and cross symbols indicate the positions of the present and  ``mean'' Sun.} %Also shown are the regression lines (solid and dashed blue) from Fig.~\ref{fig:diagr}.}
\label{fig:tr}
\end{figure}

Figure~\ref{fig:tr} shows the trajectories of the Sun over last 9000 years in the photometric variability versus chromospheric activity diagram for a few different scenarios. The upper-left panel represents  the scenario without secular variability. The $V_{\rm LT}=0.044\%$ (upper-right panel) corresponds to the secular variability reported by \cite{krivova_rec2010}.
During the last thirty years the Sun was unusually active so its mean chromospheric activity over the last 9000 years is  shifted towards lower values relative to the present position. 
The mean value of  $\log{R'_{HK}}$ equals  -4.92 without the long-term trend in the chromospheric activity, while with strong long-term trend  it equals  -4.96. 
The amplitude of the 11-year cycle in Str{\"o}mgren (b+y)/2  averaged over the last 9000 years is approximately 40\%  smaller than its present amplitude, so if  the photometric variability is not affected by a secular component then the  ``mean''  Sun is located even lower than it is now. %and the Sun is a clear outlier among Sun-like stars. 
However, a secular component in the photometric variability  shifts the ``mean''  Sun  towards larger variabilities and in combination with a displacement towards lower chromospheric activities  moves it to the regression line. 

Let us note that our basic assumption  implies that the variability of Sun-like stars is also time-dependent (see Appendix~\ref{sect:con}). In combination with the inclination effect and measurement errors this  can explain the scatter in the activity versus variability diagram, which most probably, is real.

\section{Discussion and conclusions}
%\cite{Halletal2009}  suggested that the Sun does not have unusually low variability.  
%However, we note that they used the estimate of the solar variability in the (b+y)/2 Str{\"o}mgren filters as given by  \cite{lockwoodetal2007}.  This value is approximately two times larger than the present consensus (see Fig.~\ref{fig:diagr}). Furthermore it represents the present rather than the temporal mean solar variability. 

% According to the modern consensus on the variability of solar irradiance, the present Sun is less variable than indicated by the empirical correlation between the stellar average chromospheric activity level and photometric variability. 

We have compared the available photometric records of Sun-like stars with a parameterized reconstruction of the solar irradiance over the last 9000 years. We assumed different scenarios of the solar secular variability and showed that if  the solar irradiance is not affected by  secular variations then there is a significant  disagreement between the stellar variabilities and temporary mean variability of the Sun. Part of this disagreement can be explained by the selection effect in the stellar sample \citep{Halletal2009}, relatively  small number of stars, and short time span of the stellar observations.

At the same time, we show that the presence of a secular component in the solar irradiance can increase the temporal mean of solar variability. By assuming that the temporal mean Sun obeys the activity versus regression correlation,  and employing the currently available datasets to constrain the $V_{11}$ parameter, we try to estimate the magnitude of the secular component. The calculations are presented in Appendix~\ref{sect:con}.
While the currently available stellar data yield relatively large amplitude of the secular component, careful estimation of the involved uncertainties shows that a 0.17 \% decrease of the solar 
flux in the Str{\"o}mgren (b+y)/2  filters during the Maunder minimum compared to the present value is within 95\% confidence level.  The trajectories of the Sun  in the photometric variability versus chromospheric activity diagram which correspond to this amplitude of the secular component are shown in the middle panels  of Fig.~\ref{fig:tr} (for two scenarios of the chromospheric activity variability). We note that the positions of the Sun-like stars on the variability versus activity diagram are also not necessarily fixed in time, and this can affect the variability versus activity regression. This is taken into account in Appendix~\ref{sect:con} and therefore, while the $V_{\rm LT}=0.17\%$ value is consistent with stellar data, the corresponding temporal mean position of the Sun is still below the regression line. The $V_{\rm LT}=0.17\%$ value corresponds to 1.9\,--\,2.7 W/m${^2}$  TSI change (see Appendix~\ref{sect:con}).

We  note that this number is  based solely on the empirical relationship between the stellar variability and activity  and does not depend on mechanisms of stellar and solar variability. It is given
by the present stellar sample and should be carefully re-evaluated when more stellar data become available and the low end of the activity sequence is studied.
The continuation of the stellar photometric program will thus ultimately allow one to  prove or disprove the existence of the secular variations of the solar irradiance.

\begin{acknowledgements}
The research leading to this paper was supported by the Swiss National Science Foundation under grant CRSI122-130642 (FUPSOL). We thank Phillip Judge for initiating this study and for the productive discussions, Natalie Krivova for the careful reading of the manuscript and the useful suggestions, and Wes Lockwood for providing the stellar data and helping us to understand them. {  The work has  benefited from  the COST Action ES1005 TOSCA (http://www.tosca-cost.eu)}.
\end{acknowledgements}

\bibliographystyle{aa}
%\bibliography{shapiro}

\newpage

\Online
\begin{appendix}

%\renewcommand{\thefigure}{\thechapter.\Alph{figure}}

%\clearpage

{ 
\section{Model description}\label{sect:model}
The annual solar Str{\"o}mgren  (b+y)/2  flux values $F(t)$ can be decomposed into the long-term $F^{\rm LT} (t)$ and cyclic $F^{11} (t)$ components  \citep[see, e.g.][]{tappingetal2007} 
\begin{equation}
F (t) =  F^{\rm LT} (t) + F^{11} (t). 
\label{eq:sum}
\end{equation}
While cyclic component $F^{11}(t)$ varies on the 11-year time scale and controls the activity cycle, the long-term component $F^{\rm LT}(t)$ varies on secular time scales and leads to a changeable level  of the flux during the solar activity minima. 

Most of the stars in the \citet{lockwoodetal2007} dataset are observed over an approximately 20--year period. Therefore, to compare stellar and solar variabilities we will introduce the  ${\rm RMS}_{20}$ value.
For any given year $t$, it is the RMS variability of the dataset which consists of  twenty annual flux values  (from year t-9, till year t+10).

The total variability of the solar flux can be calculated as the geometrical sum of the long-term and cyclic variabilities
\begin{equation}
{\rm RMS}_{20}(F(t))^2  = {\rm RMS}_{20}(F^{\rm LT} (t))^2  +  {\rm RMS}_{20}(F^{11} (t))^2. 
\label{eq:RMStotal}
\end{equation}

We introduce parameter $V_{11}$ as RMS variability of the annual (b+y)/2 flux values for the 1976--1996 period (solar cycles 21 and 22). 
The \cite{krivova_rec2010},  \cite{shapiro_rec}, and \cite{leanetal2005} reconstructions show the value of $V_{11}$ parameter equals  0.00019 mag, 0.00018 mag, and 0.000032 mag, respectively.  

To calculate the cyclic component backwards in time, we assume that its amplitude is proportional to the long-term solar activity $\phi (t)$. The latter  can be reconstructed over the millennia  from the cosmogenic isotope data, which gives a 22-year running mean of the modulation potential \citep{steinhilberetal2008,herbstetal2010}.  Thus, the cyclic RMS variability for any given year $t$ can be calculated as 
\begin{equation}
{\rm RMS}_{20}(F^{11} (t)) = V_{11}  \frac {\phi(t)}  {\phi(t_0)}, 
\label{eq:11final}
\end{equation}
where $\phi(t)$  is the modulation potential and  $\phi(t_0)$ is the mean value of the modulation potential for the 1976--1996 period.  The modulation potential is deduced from the composite of the neutron monitor data \citep{usoskinetal2005a} and $^{10}$Be data \citep{McCrackenetal2004,vonmoosetal2006}. It is normalised to the \cite{castagnoliandLal1980} local interstellar spectra. The homogenisation procedure is discussed in \citet{shapiro_rec} and \citet{Grec}. The average uncertainty of the modulation potential changes is $\sim$80 MeV \citep{vonmoosetal2006} and does not have a significant contribution to the temporary mean of the solar variability.

To calculate the ${\rm RMS}_{20}(F^{\rm LT} (t))$ contribution we introduce parameter $V_{\rm LT}$ as a relative change of $F^{\rm LT}$  between the present and the Maunder minimum.  The changes of the solar flux between three recent solar minima are within the uncertainties of the reconstructions and measurements of the solar irradiance \citep[see review by][]{TOSCA2012}, so the $F^{\rm LT}$ component can be approximated by a constant for the last solar cycles. Therefore, for simplicity we define $V_{\rm LT}$ as the relative change of the (b+y)/2 flux between the 1996 and the Maunder minima.  

We assume that the long-term changes of the solar irradiance are proportional to the changes of the solar activity $\phi (t)$
\begin{equation}
\frac{F^{\rm LT} (t) - F^{\rm LT} (t_0) }     {F^{\rm LT} (t_{\rm M}) - F^{\rm LT} (t_0) } = \frac{\phi (t) - \phi (t_0) }     {\phi (t_{\rm M}) - \phi (t_0) }, 
\label{eq:LT}
\end{equation}
where $t_0$ refers to  our reference period (1976--1996 average) and $t_{\rm M}$ refers to the Maunder minimum. 

Substituting $F^{\rm LT} (t_{\rm M}) - F^{\rm LT} (t_0) =   V_{\rm LT} \, F^{\rm LT} (t_0)$, one can connect the relative changes of the solar flux and activity
\begin{equation}
F^{\rm LT}(t)-F^{\rm LT}(t_0)=\frac{V_{\rm LT} \, F^{\rm LT}(t_0)}{\phi(t_{\rm M})-\phi(t_0)} \cdot  \left (  \phi(t)-\phi(t_0)      \right  ).
\label{eq:LTfinal}
\end{equation}
Equation~(\ref{eq:LTfinal}) links ${\rm RMS}_{20}(F^{\rm LT} (t))$ and $ {\rm RMS}_{20}(\phi (t)) $ via the parameter $V_{\rm LT}$. In combination with Eqs.~(\ref{eq:RMStotal}) and (\ref{eq:11final}) it allows one to reconstruct the solar variability backwards in time as a function of  $V_{11}$ and $V_{\rm LT}$ parameters.

\medskip

The approach, presented above  is based on three assumptions: 
\begin{enumerate}
\item The amplitude of the activity cycle is proportional to the long-term (i.e. cycle-averaged) solar activity.
\item The long-term changes of the solar irradiance are proportional to the long-term changes of the solar activity.
\item The long-term solar activity can be reconstructed from the cosmogenic isotope data.
\end{enumerate}

Let us note that these assumption are either employed or fulfilled because of the physical reasons  in most of the available reconstructions, independently from the proposed mechanism for the secular changes. 

To estimate the accuracy of our approach we consider three different reconstructions of the solar irradiance to the past:  \cite{krivova_rec2010} ($V_{\rm LT}$=0.044\%,  $V_{\rm 11}$=0.000019 mag), \cite{shapiro_rec}  with removed long-term trend ($V_{\rm LT}$=0\%,  $V_{\rm 11}$=0.000018 mag), and the original \cite{shapiro_rec} reconstruction $(V_{\rm LT}$=0.38\%,  $V_{\rm 11}$=0.000018 mag). They yield  0.0001 mag,  0.00009 mag, and  0.00036 mag for the mean solar RMS$_{20}$ variability over the 1620--1995 period. The same values calculated with the set of Eqs.~\ref{eq:RMStotal}--\ref{eq:LTfinal} are 0.00013 mag, 0.00012 mag, and 0.00035 mag.  One can see that our approach slightly overestimates the first two numbers. The reason for this is that to allow a homogeneous reconstruction over the millennia we use the solar modulation potential to scale the cyclic variability (see Eq.~\ref{eq:11final}) instead of the sunspot number. The annual sunspot number  is very close to zero during the Maunder minimum period, so the reconstructions predict the cessation of the cyclic variations (see Fig.~\ref{fig:comp}). At the same time the modulation potential and consequently the cyclic variability in our approach did not go to zero during the Maunder minimum \citep{McCrackenetal2004}. Such deviations are comparable to the uncertainties of the reconstructions and are not essential for our conclusions.

% Let us note that ${V}_{11}$ and $V_{\rm {lt}}$ values can not be directly compared as the first one corresponds to the RMS variability during the 20-year interval, while the second describes the total change over 300 years.

The calculation of the Ca II H and K lines, which are used to determine the solar chromospheric activity, made even more complicated by the effect of the non-local thermodynamical equilibrium \citep{ermollietal2010}.  So the mean solar chromospheric activity cannot be directly deduced from the available reconstructions of the solar irradiance over the millennia. Instead we will linearly scale it with the long-term solar activity 
\begin{equation}
R'_{\rm HK}(t)=(R'_{\rm HK})_{\rm min} + \frac{\phi(t)}{\phi(t_0)} \left (  (R'_{\rm HK})_{0}  -  (R'_{\rm HK})_{\rm min}       \right  ),
\end{equation}
where $(R'_{\rm HK})_{0}$ is the present  value of the mean solar chromospheric activity ($\log R'_{\rm HK}$=-4.895,  \citep[see][]{lockwoodetal2007} and $(R'_{\rm HK})_{\rm min} $ is the value of the mean chromospheric activity corresponding to $\phi=0$.

In Fig.~\ref{fig:tr} we consider two extreme scenarios (see Sect.~\ref{sect:historical}), one without the secular component in the solar chromospheric activity, and another with a strong secular component.
Without the secular component the changes of the mean solar chromospheric activity are only caused by the variable amplitude of the solar activity cycle and the $(R'_{\rm HK})_{\rm min}$ value corresponds to the present value of the solar activity at minimum conditions ($\log R'_{\rm HK}$=-4.98; \citep[see][]{judgesaar2007}. To describe the case of a strong secular component in the chromospheric activity we put the $(R'_{\rm HK})_{\rm min}$ value equals to the boundary value from \citet{saar2006} ($\log{R'_{HK}}=-5.08$).

}

%\begin{equation}
%\frac{F^{\rm LT}(t_1)-F^{\rm LT}(t_2)}{F^{\rm LT}(t_1)} = V_{\rm LT} \frac{\phi (t_1)}{\phi(t_M)-\phi(t_0)}  \cdot \frac{\phi(t_1)-\phi(t_2)}{\phi(t_1)}
%\end{equation}

%\begin{equation}
%{\rm RMS}_{20}(F^{\rm LT} (t)) = V_{\rm LT} \frac{\phi (t)}{\phi(t_M)-\phi(t_0)} \cdot {\rm RMS}_{20}(\phi (t)) 
%\end{equation}

\section{Possible constraints on the historical solar variability}\label{sect:con}
\begin{figure*}
\resizebox{0.9\hsize}{!}{\includegraphics{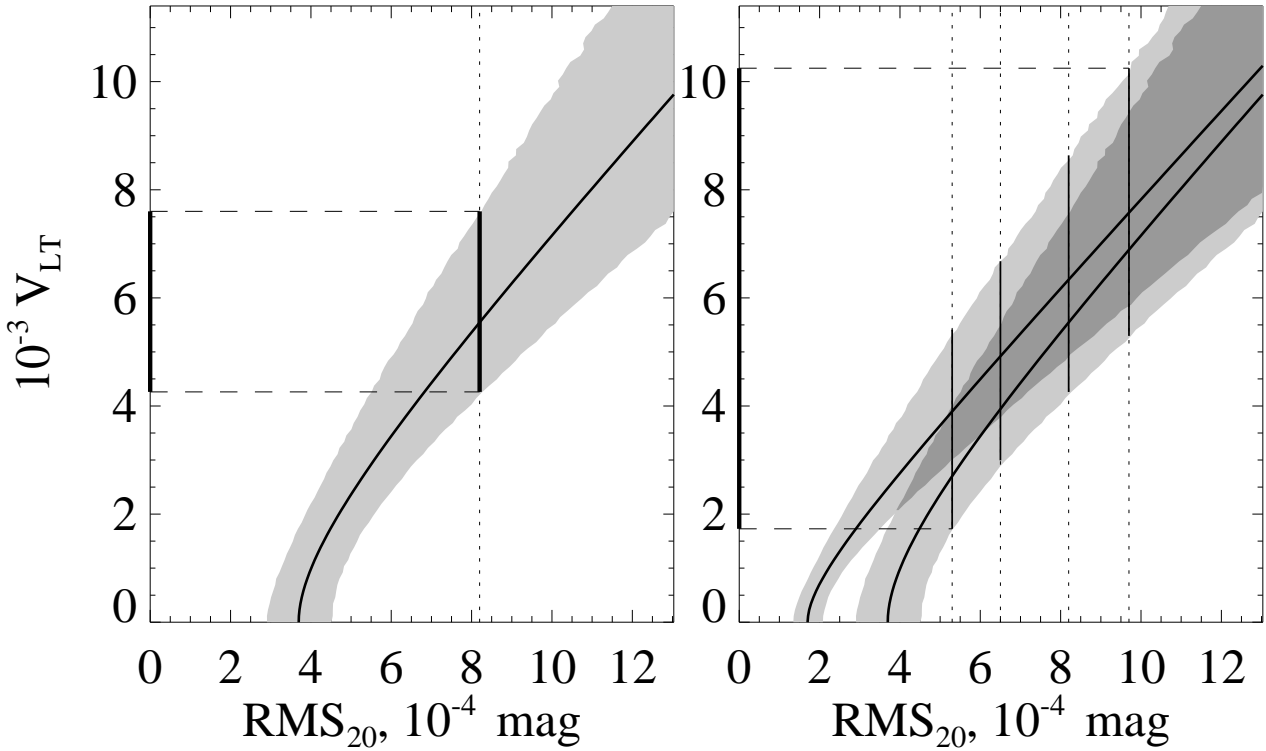}}
\caption{%Left panel: Comparison of the theoretical tail distributions of the solar variability (which show the probability that the variability is above a particular level) with the tail distribution of the stellar variabilities from \citet{lockwoodetal2007},  scaled to the solar chromospheric activity level (star symbols). Theoretical distributions are plotted for four combinations of the $V_{\rm {LT}}$ and  ${V}_{11}$ pairs: two   $V_{\rm {LT}}$ values:  $3.8\cdot10^{-3}$ (thin curves) and  $6.4\cdot10^{-3}$  (thick curves), and two ${V}_{11}$ values:  0 (solid curves) and $4 \cdot 10^{-4}$ (dashed curves).  
Left panel: The dependency of the $V_{\rm LT}$ parameter on the solar variability, retrieved from the stellar data. The shaded area indicates the uncertainty (2$\sigma$) due to the limited number of stars.
The ${V}_{11}$ parameter is set to 0.00044 mag.  The dashed line corresponds to the expected mean solar variability RMS$_{20}=0.00082$. The projection onto  $V_{\rm {LT}}$-axis yields the 95\% interval for expected long-term variability of the Sun: $0.43\%<V_{\rm LT} < 0.76\%$. 
Right panel: The same as the left panel, but  the contour calculated with  ${V}_{11}$  = 0.0002 mag  is added. The four dashed lines correspond to four values of expected mean solar variability RMS$_{20}$:  0.00097 mag (no  chromospheric secular changes, regression with solid line from Fig.~\ref{fig:diagr}), 0.00082 mag (strong chromospheric secular changes, regression with solid line), 0.00065 mag (no  chromospheric secular changes, regression with dashed line), 0.00053 mag (strong chromospheric secular changes, regression with dashed line). The projection onto the $V_{\rm {LT}}$-axis yields the interval $0.17\%<V_{\rm LT}<1.2\%$.  }
\label{fig:pdf}
\end{figure*}

\begin{figure*}
\resizebox{\hsize}{!}{\includegraphics{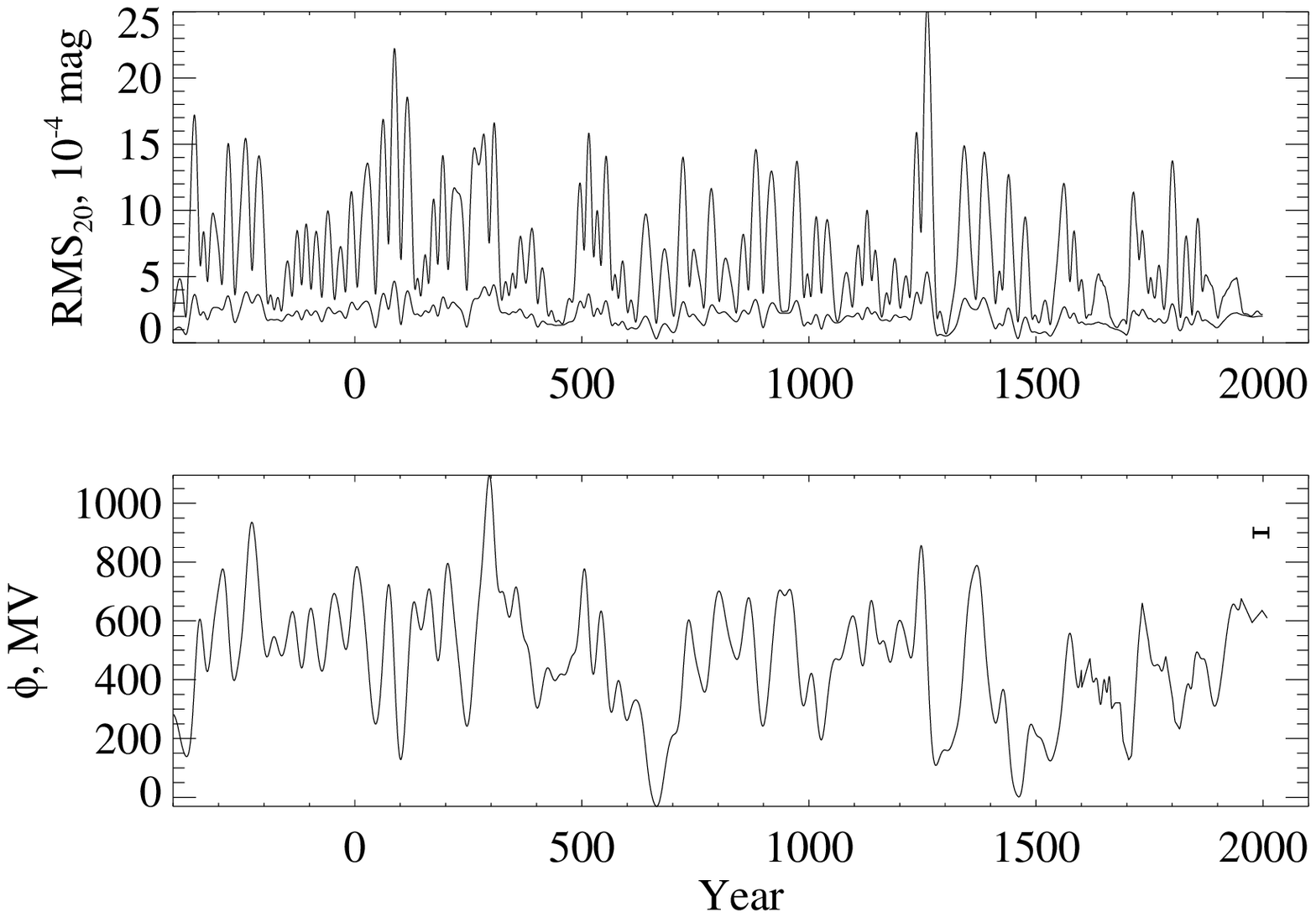}}
\caption{Upper panel: RMS$_{20}$ back  to 400 BC,  adopting  $V_{\rm {LT}}=0.001$ (lower curve) and  $V_{\rm {LT}}=0.005$ (upper curve). ${V}_{11}$ is set to 0.0002 mag. Lower panel: Modulation potential back to 400 BC. The interval in the right upper corner denotes the satellite era of continuous solar irradiance measurements.}
\label{fig:rec2400}
\end{figure*}

The amplitude of the secular solar photometric variability can be constrained by demanding  the temporal mean solar variability be in agreement with the variability, indicated by the stellar data. The latter depends on the version of the variability versus activity  regression and the scenario of the chromospheric activity behaviour. To investigate { different} cases in Fig.~\ref{fig:pdf}  we plot the  long-term solar variability $V_{\rm { LT}}$  as a function of the mean solar variability (RMS$_{20}$) given by the stellar data, for two values of the 11-year variability ${V}_{11}$. The inclination effect was calculated according to \cite{knaacketal2003}.

{Let us note, however, that the positions of the Sun-like stars in  Fig.~\ref{fig:tr} are also not necessarily fixed in time. This can affect the coefficients of the regression line and accordingly the estimated variability of the Sun. For example it is possible that, by coincidence,  most of the stars among the twenty-one used for our analysis represent the periods of the relatively low  (or high) variability. This will lead to the shifted  position of the regression line and consequently to the deviations in the $V_{\rm {lt}}$  parameter determined using this limited selection of stars.

If the behaviour of the stellar variabilities after the regression to the solar level of the chromospheric activity is identical to the solar variability, i.e. the solar and regressed stellar variabilities can be parameterised employing the approach described in Sect.~\ref{sect:historical} and Appendix~\ref{sect:model}, then the resulting uncertainty can be found by performing  the Monte Carlo simulation. To do this we considered a large number of sets consisting of 21 stars. Every star in these sets represents the Sun at some random twenty-year interval period (chosen from the full 9000-year dataset) and with a random angle between the stellar rotation axes and the direction to the observer. The RMS$_{20}$ variability of these stars is calculated, employing the approach described in Sect.~\ref{sect:historical} and Appendix \ref{sect:model} and the \citet{knaacketal2003}  dependency of the 11-year variability on the angle  between the solar rotation axes and the direction to the observer. For every set we calculated the mean RMS$_{20}$  variability of the stars in the set, so for every pair of $V_{\rm LT}$ and $V_{11}$ parameters we have a distribution of mean RMS$_{20}$ variabilities. This allows us to calculate the range of $V_{\rm LT}$ and $V_{11}$  parameters which can lead to the present value of the linear coefficients of the activity versus variability regression line. } In Fig.~\ref{fig:pdf} we mark the resulting $2 \sigma$ uncertainty in $V_{\rm { LT}}$  parameter with a shaded area.

The mean solar $\log{R'_{HK}}$  index  in the case of the strong secular component in chromospheric activity is -4.96 (see Fig.~\ref{fig:tr}). Using the solid blue line in Fig.~\ref{fig:diagr} one can find that this index corresponds to the $8.2 \cdot 10^{-4}$ RMS variability. This value is inserted in the left panel of Fig.~\ref{fig:pdf} and one can see that it yields $V_{\rm { LT}} = (0.43 \% - 0.76 \%) $. 

In the right panel of  Fig.~\ref{fig:pdf}  we also plotted the contour with ${V}_{11}  = 0.0002$ mag and indicated four different values of the expected solar variability, corresponding to different treatments of the chromospheric secular changes and two  regressions from Fig.~\ref{fig:diagr}. One can see that the minimum  value of   $V_{\rm { LT}}$ equals  $0.17 \%$. 

The parameter $V_{\rm { LT}}$ characterises the long-term variability as it is observed in the Str{\"o}mgren b and y filters. Its connection with the  variability at another spectral domain (or TSI) depends on the mechanism of the secular changes. For example a $V_{\rm { LT}}$ equal to $0.17 \%$ leads to a  2.7 W/m${^2}$ TSI change  between the Maunder and the last  solar minima, adopting the model of \cite{shapiro_rec}, and to 1.9 W/m${^2}$, assuming that it is caused by the variations of the solar effective temperature.

%One can see that even relatively small values of the secular solar variability can force the Sun to obey the empirical relationship between the  photometric variability and mean chromospheric activity. 

%\begin{figure}
%\resizebox{0.8\hsize}{!}{\includegraphics{stellar5.eps}}
%\caption{Probability distribution functions for the minimum, width and mean of the random set of 21 stars. The variabilities of the stars obey to the distribution calculated with   two $V_{\rm {lt}}$ values:  $3.8\cdot10^{-3}$ (thin curves), $6.3\cdot10^{-3}$  (thick curves), and two ${V}_{11}$ values:  $1 \cdot 10^{-4}$  (solid curves) and $4 \cdot 10^{-4}$ (dashed curves).
%Dash dot lines indicate the values determined from the stellar set. }
%\label{fig:lim}
%\end{figure}

%\begin{figure}
%\plottwo{f2.eps}{f2_color.eps}
%\caption{A panel taken from Figure 2 of \citet{rudnick03}. 
%See the electronic edition of the Journal for a color version 
%of this figure.\label{fig2}}
%\end{figure}

\end{appendix}
\end{document}